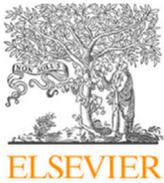
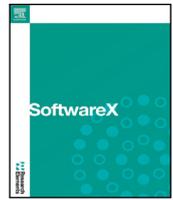

Original Software Publication

# PATSMA: Parameter Auto-tuning for Shared Memory Algorithms

Joao B. Fernandes [a],[*], Felipe H. Santos-da-Silva [b], Tiago Barros [a], Italo A.S. Assis [b], Samuel Xavier-de-Souza [a]

[a] *Department of Computer Engineering and Automation, UFRN, Natal, Brazil*
[b] *Department of Engineering and Technology, UFERSA, Pau dos Ferros, Brazil*

A B S T R A C T

Programs with high levels of complexity often face challenges in adjusting execution parameters, particularly when the ideal value for these parameters may change based on the execution context. These dynamic parameters significantly impact the program's performance. For instance, ideal parallel loop scheduling may vary depending on factors like the execution environment, program input, or the choice of compiler. Given the expensive nature of testing each case individually, one viable solution is to automate parameter adjustments using optimization methods. This article introduces PATSMA, a parameter auto-tuning tool that leverages Coupled Simulated Annealing (CSA) and Nelder–Mead (NM) optimization methods to fine-tune existing parameters in an iterative application. We demonstrate how auto-tuning can contribute to the real-time optimization of parallel algorithms designed for shared memory systems. PATSMA is a C++ library readily available under the MIT license.

| | |
|---|---|
| Current code version | 1.0 |
| Permanent link to code/repository used for this code version | https://github.com/ElsevierSoftwareX/SOFTX-D-24-00176 |
| Permanent link to Reproducible Capsule | https://github.com/lapps-ufrn/PATSMA |
| Legal Code License | Legal Code License |
| Code versioning system used | git v2.27.0 |
| | gnu 11.2.0 |
| | cmake 3.21.3 |
| Software code languages, tools, and services used | C++ |
| Compilation requirements, operating environments & dependencies | Linux |
| If available Link to developer documentation/manual | https://lappsufrn.gitlab.io/patsma |
| Support email for questions | joao.batista.fernandes.094@ufrn.edu.br |

## 1. Motivation and significance

When working with parallel programming, one of the most significant challenges is appropriately adjusting parameters, especially considering that certain parameters depend on external variables such as CPU, memory, and the number of threads. Moreover, there often is not a clear mathematical relationship between an application's execution time and the tuning of these parameters. Consequently, exploring various parameter variations becomes necessary to find an optimal configuration. In this context, auto-tuning proves invaluable. It addresses the complexity of applications by accounting for factors like memory availability, workload significance, control flow deviations, and idle cores. These factors may fluctuate during runtime, making it crucial to have an auto-tuning mechanism that can dynamically adapt to such variations more precisely.

PATSMA (Parameter Auto Tuning for Shared Memory Algorithms) is a C++ library for parameter optimization based on various optimization methods, including Coupled Simulated Annealing (CSA) [1] and Nelder–Mead (NM) [2]. We design the library to be easily extendable to accommodate other optimization techniques. PATSMA primarily focuses on optimizing the execution time of a given code segment, but it also allows for utilizing other program variables as optimization parameters. This versatility makes the library adaptable to a range of use cases. PATSMA's functionality is integrated into the code by including specific functions and adjustments to user applications.

Several authors have proposed auto-tuning techniques. Andreolli et al. [3,4] introduce a self-tuning framework for seismic applications, wherein each set of parameters selected by a genetic algorithm is compiled and executed. Some authors [5–7] present generic auto-tuning for several applications, such as matrix multiplication, fast Fourier transform, and 2D convolution. The mentioned works [3–7] perform auto-tuning before runtime differently of PATSMA. Rash et al. [8] present ATF (Automatic Tuning Framework), a comprehensive approach to auto-tuning programs with interdependent tuning parameters. Unlike our auto-tuning approach, where the search space is defined in advance without needing a generation step, ATF analyzes previous constraints to generate a search space. Petrovič and Filipovič [9] introduce an on-line auto-tuning named KTT (Kernel Tuning Toolkit). However, unlike PATSMA, it does not apply any optimization methods.

PATSMA has been employed in various research projects. Barros et al. [10] and Fernandes et al. [11] optimized a 3D finite difference method (FDM) algorithm, while Assis et al. [12] and Fernandes et al. [13] applied it to a 3D RTM algorithm. These studies employed PATSMA to adjust parallel loop scheduling and, thus, reduce program execution time.

## 2. Software description

PATSMA library is an open-source project with an MIT license, developed in C++ and incorporating OpenMP parallelism. Its implementation establishes an interface for extracting information from the target method and employing optimization algorithms to determine the best responses for execution. The library includes two implemented optimization algorithms: Coupled Simulated Annealing (CSA) [1] and Nelder–Mead (NM) [2]. However, as described in Section 2.2, other algorithms can be incorporated.

### 2.1. PATSMA overview

CSA is an algorithm derived from Simulated Annealing (SA) [14] and is characterized by orchestrating the execution of multiple SA optimizers. It extracts information through a coupled term, facilitating the diversification of these optimizers between global and local searches. Consequently, CSA exhibits considerable flexibility in blending refined searches with escapes from local minima, establishing it as the primary algorithm within our library. The other implemented method, NM, is a well-known optimizer recognized for its more direct approach, often delivering quicker results. However, NM is prone to becoming trapped in local minima. Therefore, it is better suited for simpler problems.

PATSMA can operate in two execution modes: *Single Iteration* and *Entire Execution*. In the *Single Iteration* mode, depicted in Fig. 1(a), the target method slated for adjustment is integrated with the PATSMA execution function. Consequently, the optimization process and target method execution unfold in tandem. If the number of target method iterations is sufficient for the optimization to complete, PATSMA will bypass the optimization and run the remaining target iterations using the computed final solution. In this mode, users experience minimal execution overhead, with the only additional computation being that of the optimizer. Consequently, the higher the cost of the target method, the lower the proportion of overhead.

While the *Single Iteration* mode may seem ideal, using it is not always possible. Sometimes, the advancement in the execution of the target method can impede optimization. For example, different target iterations may measure varying costs for the same solution, especially when this cost is related to a performance parameter. Hence, we introduced the *Entire Execution* mode, in which all optimizer executions are separate from the target itself. To attain this objective, the user must invoke PATSMA outside the target loop, utilizing a replica of the target method and identical parameters. PATSMA will then execute using this replica. The user must ensure that any parameter utilized during the replica execution (excluding the adjusted parameter) will impact the original execution. This approach results in additional execution of target iterations, leading to a noticeable surge in overhead, which may reduce the overall benefit.

In both cases, the *Numerical Optimizer* in Fig. 1 represents the PATSMA method responsible for calculating the new *Candidate Solution* at each auto-tuning iteration to be tested by the target method. For each *Candidate Solution* will be calculated a respective *Cost*. For this, PATSMA measures the target method runtime using the *Candidate Solution*. After executing the target method iterations, PATSMA sends the runtime as a *Cost* to the *Numerical Optimizer*. However, both the *Single Iteration* and *Entire Execution* functions also incorporate a *Runtime* mode, wherein the *Start Measure* and *End Measure* methods are responsible for measuring the execution time of the target. Consequently, this time is then returned to the *Numerical Optimizer* as the corresponding *Cost*, relieving the application of this responsibility. However, using only other attributes returned by the target method as a *Cost* is possible.

### 2.2. Incorporating optimization algorithms

We specify the optimizer employed in the *Numerical Optimizer* during library startup, with the default being the CSA. Although PATSMA currently only implements CSA and Nelder–Mead optimization algorithms, other optimization methods can be incorporated as a new class. This new optimizer class needs to extend the *NumericalOptimizer* interface outlined in Algorithm 1. Four methods must be developed for this extension, as indicated in lines 6 to 9 of Algorithm 1. Additionally, as depicted in lines 10 and 11 of Algorithm 1, two other methods are optional.

**Algorithm 1** Numerical optimizer interface.

```
 1 class NumericalOptimizer {
 2 public:
 3   NumericalOptimizer(){};
 4   virtual ~NumericalOptimizer() {}
 5
 6   virtual double* run(double cost) = 0;
 7   virtual int getNumPoints() const = 0;
 8   virtual int getDimension() const = 0;
 9   virtual bool isEnd() const = 0;
10   virtual void reset(int level) {};
11   virtual void print() const {}
12 };
```

The methods getNumPoints(), getDimension(), and isEnd() are functions that return necessary data for execution. They provide the number of solutions (point) that the optimization algorithm returns, the number of dimensions the solutions have, and whether the optimization has ended, respectively. On the other hand, reset(int level) is an option to reset the optimization and allows the definition of levels for the reset. Generally, a zero level corresponds to a lighter reset that retains most of the information, such as the solutions found, while higher levels, up to a maximum, result in a complete reset of the optimization. Additionally, print() is another optional method that allows printing debug or verbose information from your optimizer.

Finally, run(**double** cost) is the primary function. This method receives the cost and returns the candidate solution. The cost pertains solely to the last solution returned by the preceding run call. Therefore, the initial run call need not receive a consistent cost value as it will not be employed by the optimization algorithm. Furthermore, upon completing the optimization process, the run function will provide the final solution, which does not require further testing. Generally, the run method must operate in stages. Each time the method calculates a new candidate solution, it should return it. This behavior is necessary because the library covers various cases, including situations where the target method cannot express the cost function as an equation or a method within the program, such as measuring the execution time of a piece of code. Beyond that, this approach minimizes the need to modify the target application code and enables optimization during program execution. Instead of defining a cost function and passing it to the run method.

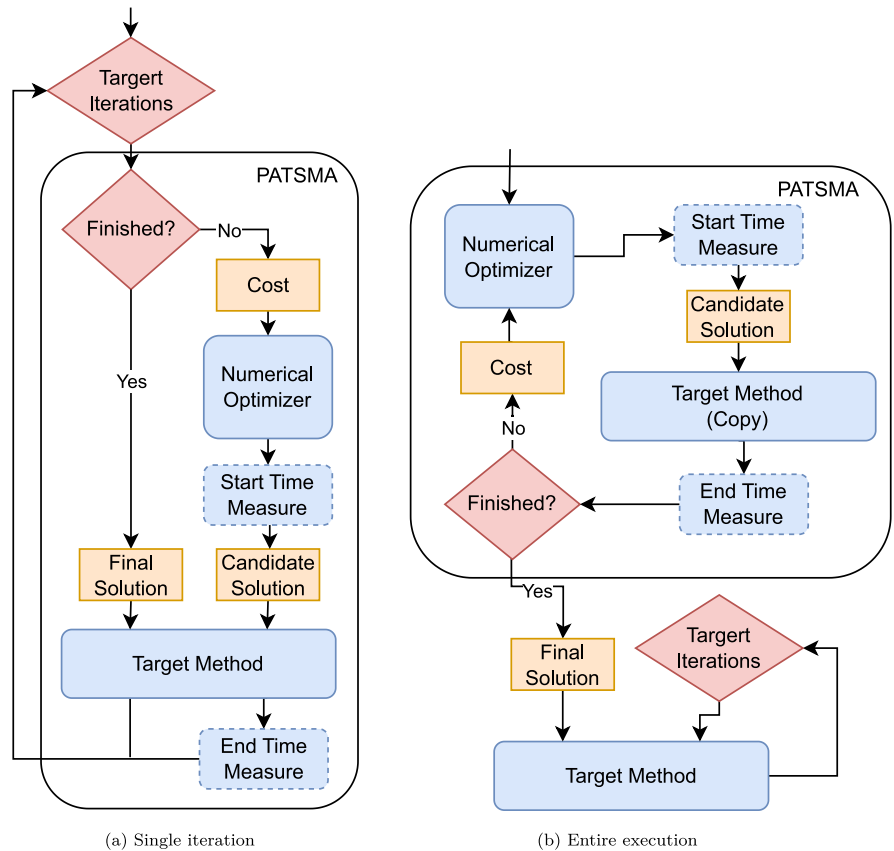

**Fig. 1.** PATSMA library when it is (a) coupled with the method, executing a single auto-tuning iteration each time, and when it is (b) decoupled, doing the entire auto-tuning execution before the method. Dashed methods are eventual.

*2.3. Library setup*

To use the library, developers must carefully assess which parameters of their target application can be adjusted and specify their optimization goals accordingly. While any parameter can be adjusted, those related to performance are particularly well-suited for this library. These may include block size (or loop granularity), number of threads, and clock frequency. By default, the library is set up to minimize the execution time of a given piece of code or function, which is especially useful for complex algorithms. However, developers can also optimize other parameters necessary for their specific goals by passing them as parameters to the library, such as energy consumption.

Before initializing the library, the user must first import it using the following C++ include:

```
1 #include "Autotuning.hpp"
```

By importing *auto-tuning*, all other necessary files for the library will be automatically imported, including CSA and Nelder–Mead. **auto**—tuning .hpp contains the *auto-tuning* class, which includes the variables and methods responsible for creating a management interface between optimization algorithms and the application. This class can be initialized in two ways. In both initialization methods, the user has to define the values of *min*, *max*, and *ignore*. They are essential parameters of the auto-tuning class and are the minimum solution value, the maximum solution value, and the number of target iterations ignored in each solution cost calculation. When an application uses performance as a cost function, it may encounter target iterations that interfere with the calculation of the current solution. In such cases, the *ignore* parameter can be utilized to remove these target iterations and allow the execution to stabilize. This stabilization process generally requires one additional iteration of the target.

In Algorithm 2, Line 4 corresponds to the constructor function that utilizes the default numerical optimizer (CSA). Within this function, the specific parameters *dim*, *num_opt*, and *max_iter* define aspects of CSA's optimization process. These parameters indicate the problem's dimensionality, the number of optimizing entities (or solutions calculated for each), and the maximum number of iterations CSA can perform. It is important to note that the *max_iter* differs from the number of cost function evaluations (*num_eval*). Nonetheless, when using CSA, there is a functional relationship between the two, which can be expressed by the following equation:

$$num\_eval = max\_iter * (ignore + 1) * num\_opt \qquad (1)$$

**Algorithm 2** Auto-tuning class constructors.

```
1 class Autotuning{
2     [...]
3   public:
4     autotuning(double min, double max, int ignore, int
            dim, int num_opt, int max_iter);
5     autotuning(double min, double max, int ignore,
            NumericalOptimizer *optimizer);
6     [...]
7 }
```

The alternative constructor option is depicted in Line 5 of Algorithm 2, where we can specify the optimization algorithm we wish to employ.

We can initialize the CSA algorithm using the constructor CSA(**int** dim, **int** num_opt, **int** max_iter), with its parameters remaining consistent with those previously introduced in the auto-tuning constructor. Conversely, we can utilize the constructor NelderMead(**int** dim, **double** error, **int** max_iter = 0) for the Nelder–Mead algorithm. Here, *dim* continues to denote the dimension of the problem, while *error* and *max_iter* serve as the stopping criteria for minimum error and the maximum number of iterations, respectively. The parameter *max_iter* is optional, and Eq. (2) described its relationship with the number of evaluations for NelderMead.

$$num\_eval = max\_iter * (ignore + 1) \quad (2)$$

*2.4. PATSMA execution*

To utilize PATSMA, we can make use of the base methods (Line 5 to Line 8 of Algorithm 3) and pre-programmed methods (Line 10 to Line 16 of Algorithm 3). As PATSMA was designed to minimize execution time, the user can specify the code section they wish to optimize and use the start(Point *point) and end() methods to set the section's boundaries. Every time the application executes the code segment, PATSMA performs an auto-tuning iteration. Section 3 shows an example of this process.

Furthermore, users can set their preferred cost optimization by utilizing the exec(Point *point, **double** cost) method. This method allows the target application to generate its own cost value and receive the next candidate solution (point) from the PATSMA in response. How the cost is defined by the application, there are no code boundaries, so it can be used wherever the cost is generated. In this way, the PATSMA will work as a simple optimization algorithm. It is worth noting that the cost value is always associated with the last returned solution, following the same logic outlined in the run method discussed in Section 2.2. Although the dimension of the solution is predetermined in the class constructor method, users must indicate the point type in the execution methods if different from the *int* type, as exemplified in the following exec line code with *double* type

```
1 exec<double>(point, cost);
```

It is crucial to remember that type is restricted to integer or floating-point arithmetic types.

Pre-programmed methods streamline programming processes by allowing developers to quickly insert code snippets into their applications. However, ensuring these functions exhibit two key characteristics is essential. Firstly, the initial variable must serve as both an input and output for the calculated point when using any of these methods. Secondly, methods lacking the *Runtime* suffix require the function to return the cost associated with the point. This suffix refers to the execution time as a cost already measured within the method and does not require the user to pass this value explicitly.

Algorithm 3 includes methods prefixed with *entire* on Lines 10 and 12, which execute the complete auto-tuning process and return the result point promptly. These methods are recommended for cases where the adjustment must be executed before entering a loop. Conversely, methods prefixed with *single* on Lines 14 and 16 perform only one iteration of the auto-tuning each time the program passes through them.

## 3. Illustrative examples

Our research showcases the application of the Red-Black (RB) method in parallelizing the iterative Gauss–Seidel method as a code example. This method utilizes element coloring to divide the problem into alternating black and red subdomains. By updating black elements first and then red elements, the RB method allows for parallel updates for each color, as elements of different colors have no dependencies on each other. The Gauss–Seidel iteration continues until the solution reaches a satisfactory convergence.

**Algorithm 3** Execution methods.

```
1  class Autotuning{
2      [...]
3   public:
4      template <typename Point>
5      void start(Point *point);
6      void end();
7      template <typename Point>
8      void exec(Point *point, double cost);
9      template <typename Point = int, typename Func,
            typename... Args>
10     void entireExecRuntime(Func function, Point *point
            , Args... args);
11     template <typename Point = int, typename Func,
            typename... Args>
12     void entireExec(Func function, Point *point, Args
            ... args);
13     template <typename Point = int, typename Func,
            typename... Args>
14     void singleExecRuntime(Func function, Point *point
            , Args... args);
15     template <typename Point = int, typename Func,
            typename... Args>
16     void singleExec(Func function, Point *point, Args
            ... args);
17     [...]
18 }
```

The Algorithm 4 showcases implementing RB Gauss–Seidel's matrix _calculation function in C++. The function consists of two *for* loops, which calculate red and black elements of the matrix, respectively. To optimize performance, these loops are parallelized with OpenMP, utilizing a scheduling clause of schedule(dynamic, chunk). However, determining the optimal value for the chunk can be difficult, as it depends on various factors such as system specifications, input data size, and type. That is where PATSMA comes in handy, an invaluable tool for automatically adjusting this value.

**Algorithm 4** Red-black algorithm without PATSMA

```
1  double matrix_calculation(double **A, int n) {
2      // ... (Initialization code)
3
4      #pragma omp parallel private(tmp, i, j) {
5      #pragma omp for reduction(+ : diff) schedule(
            dynamic, chunk)
6      for (i = 1; i <= n; ++i) {
7          for (j = 1; j <= n; ++j) {
8              // ... (Update black elements)
9          }
10     }
11     #pragma omp for reduction(+ : diff) schedule(
            dynamic, chunk)
12     for (i = 1; i <= n; ++i) {
13         for (j = 1; j <= n; ++j) {
14             // ... (Update red elements)
15         }
16     }
17     return diff;
18 }
```

A minor modification is necessary to optimize the application method outlined in Algorithm 4 using PATSMA. Simply append the parameter to be adjusted at the end of the method signature. The code snippet below shows an example using a pointer for the parameter chunk, as it may consist of multiple dimensions in some cases:

```
double matrix_calculation(double **A, int n, int *chunk) {
    [...] }
```

Since the application involves two nested loops with two distinct values for the chunk variable, there are two potential approaches. One option is to use a single value for both loops:

```
double matrix_calculation(double **A, int n, int *chunk) {
    [...]
    #pragma omp for reduction(+:diff) schedule(dynamic,
        chunk[0])
    [...]
    #pragma omp for reduction(+:diff) schedule(dynamic,
        chunk[0])
    [...]
}
```

Alternatively, the values for the chunk variable can be calculated separately and used as follows:

```
double matrix_calculation(double **A, int n, int *chunk) {
    [...]
    #pragma omp for reduction(+:diff) schedule(dynamic,
        chunk[0])
    [...]
    #pragma omp for reduction(+:diff) schedule(dynamic,
        chunk[1])
    [...]
}
```

With this approach, we can employ the PATSMA execution methods outlined in Algorithm 3. In the first version, we utilize the entireExecRuntime method, which adjusts the *chunk* based on the execution time that each value generates as a cost. The entireExecRuntime is placed outside the Gauss–Seidel iteration loop as illustrated in Algorithm 5 because it executes *n_iter* auto-tuning iterations and returns the final value of *chunk*.

**Algorithm 5** RB Gauss-Seidel using PATSMA to optimize load balancing with the function entireExecRuntime.

```
 1 #include "Autotuning.hpp"
 2
 3 void solve_parallel(double **A, int n){
 4
 5 // ... (Gauss-Seidel initialization code)
 6 // ... (Auto-tuning parameters definitions)
 7
 8 autotuning *at = new autotuning(min,max,ignore,dim,
       n_opt,n_iter);
 9 at->entireExecRuntime(matrix_calculation, chunk, A, N
       -1);
10
11 for (for_iters = 1; for_iters < max_iter; ++for_iters)
       {
12    diff = 0;
13    diff = matrix_calculation(A, N - 1, chunk);
14    iters++;
15    [...]
16  }
17 }
```

Another version of the same algorithm is demonstrated in Algorithm 6. In this case, we utilize the singleExecRuntime execution function, where each function call executes only one iteration of the matrix_calculation application. This way, the optimization will occur together with the natural execution of the application until the optimization concludes. When the optimization finishes, the singleExecRuntime continues being called, but only the application is executed using the final solution to the chunk parameter. Therefore, unlike Algorithm 5, the PATSMA function is placed within the execution loop. This minimizes overhead, as it does not create extra matrix_calculation iterations, restricting itself only to the additional computation of the optimization method itself.

**Algorithm 6** RB Gauss-Seidel using PATSMA to optimize load balancing with the singleExecRuntime function.

```
 1 #include "Autotuning.hpp"
 2
 3 void solve_parallel(double **A, int n){
 4 // ... (Gauss-Seidel initialization code)
 5 // ... (Auto-tuning parameters definitions)
 6 // ... (Auto-tuning initialization code)
 7
 8 for (for_iters = 1; for_iters < max_iter; ++for_iters)
       {
 9    diff = 0;
10    diff = at->singleExecRuntime(matrix_calculation,
          chunk, A, N-1);
11    iters++;
12    // ... (Test and print code)
13  }
14 }
```

## 4. Impact

The PATSMA library, incorporated in this work, is an open-source solution designed to simplify parameter tuning. Its code is implemented in C++, utilizing optimization algorithms such as CSA and Nelder–Mead to enhance parameter values. It affords users the flexibility to implement and employ their optimizer. The primary objective of PATSMA is to automate the tuning of performance parameters that occur in runtime. These parameters are known for their sensitivity to the execution environment, with various hardware and software elements influencing performance, such as the number of cores, cache size, memory speed, or control flow structure like *if else*.

The environment particularly impacts load balancing and task distribution. In the context of OpenMP, a parallel programming tool for shared memory environments, the size of the *chunk* is critical for loop scheduling. Without the auto-tuning library, users are constrained to selecting an arbitrary value, hoping that it will function adequately in any given execution environment, or resorting to a time-consuming trial-and-error process. In our prior work, we illustrated the effectiveness of the PATSMA library in automatically adjusting the size of the *chunk* [10–13].

## CRediT authorship contribution statement

**Joao B. Fernandes:** Writing – original draft, Visualization, Validation, Software, Methodology, Investigation, Formal analysis, Data curation, Conceptualization. **Felipe H. Santos-da-Silva:** Writing – original draft, Software, Investigation. **Tiago Barros:** Writing – review & editing. **Italo A.S. Assis:** Writing – review & editing, Supervision. **Samuel Xavier-de-Souza:** Resources, Project administration, Funding acquisition.

## Declaration of competing interest

The authors declare the following financial interests/personal relationships which may be considered as potential competing interests: Joao Batista Fernandes reports financial support and article publishing charges were provided by Shell Brazil Oil. Tiago Barros reports financial support and article publishing charges were provided by Shell Brazil Oil. Samuel Xavier-de-Souza reports financial support and article publishing charges were provided by Shell Brazil Oil. Italo A. S. Assis reports financial support and article publishing charges were provided by Shell Brazil Oil. If there are other authors, they declare that they have no known competing financial interests or personal relationships that could have appeared to influence the work reported in this paper.

**Data availability**

Data will be made available on request.

**Acknowledgments**

The authors gratefully acknowledge support from Shell Brazil through the project "*Novas metodologias computacionalmente escaláveis para sísmica 4D orientado ao alvo em reservatórios do pré-sal*" at the Universidade Federal do Rio Grande do Norte (UFRN) and the strategic importance of the support given by ANP through the R&D levy regulation. Felipe Silva has been an undergraduate researcher through the PIVIC and PIVIC-Af programs at Universidade Federal Rural do Semi-Árido (UFERSA). This work was partly supported by Oracle Cloud credits and related resources the Oracle for Research program provided. The authors also acknowledge the National Laboratory for Scientific Computing (LNCC/MCTI, Brazil) and the High-Performance Computing Center at UFRN (NPAD/UFRN) for providing HPC resources of the SDumont and NPAD supercomputers.